\documentclass[sn-mathphys,Numbered]{sn-jnl}

\usepackage{graphicx}
\usepackage{color}
\usepackage{amssymb}
\usepackage{amsmath}
\usepackage{array}
\usepackage{mathtools}
\usepackage{placeins}
\usepackage{float}
\usepackage{tabularx}
\usepackage{adjustbox}
\usepackage{xurl}
\usepackage{hyperref}
\usepackage{natbib}
\usepackage[normalem]{ulem}
\usepackage{afterpage}
\usepackage{float}
\usepackage{ulem}
\usepackage{tensor}
\usepackage[T1]{fontenc}
\usepackage[utf8]{inputenc}
\usepackage{multirow}%
\usepackage{amsthm}%
\usepackage{mathrsfs}%
\usepackage[title]{appendix}%
\usepackage{xcolor}%
\usepackage{textcomp}%
\usepackage{manyfoot}%
\usepackage{booktabs}%
\usepackage{algorithm}%
\usepackage{algorithmicx}%
\usepackage{algpseudocode}%
\usepackage{listings}%

\newcommand{\bn}{{\mathbf n}}
\newcommand{\bv}{{\mathbf v}}

\newcommand{\de}{\delta}
\newcommand{\De}{\Delta}

\newcommand{\be}{\begin{equation}}
\newcommand{\ee}{\end{equation}}

\newcommand{\bea}{\begin{eqnarray}}
\newcommand{\eea}{\end{eqnarray}}
\newcommand{\bean}{\begin{eqnarray*}}
\newcommand{\eean}{\end{eqnarray*}}

\newcommand{\dd}[0]{\mathrm{d}}

\newcommand{\alphaT}[0]{\alpha_{\mathrm{T}}}
\newcommand{\alphaL}[0]{\alpha_{\mathrm{L}}}

\newcommand\spart{\;\raise1.0pt\hbox{$/$}\hskip-6pt\partial}
\definecolor{dred}{rgb}{0.8,0,0.1}
\definecolor{orange}{rgb}{1,0.5,0}

\begin{document}

\title[Invariance of cosmological number counts under disformal transformations]{Invariance of cosmological number counts under disformal transformations}
\author*[1]{\fnm{Basundhara} \sur{Ghosh}}\email{basundharag@iisc.ac.in}

\author[2]{\fnm{J\'er\'emie} \sur{Francfort}}\email{jeremie.francfort@unige.ch}

\author[1]{\fnm{Rajeev Kumar} \sur{Jain}}\email{rkjain@iisc.ac.in}

\affil*[1]{\orgdiv{Department of Physics}, \orgname{Indian Institute of Science}, \orgaddress{\street{CV Raman Road}, \city{Bengaluru}, \postcode{560012}, \state{Karnataka}, \country{India}}}

\affil[2]{\orgdiv{Department of Theoretical Physics and Center for Astroparticle Physics}, \orgname{University of Geneva}, \orgaddress{\street{24 quai Ernest  Ansermet}, \city{Geneva}, \postcode{1211},\country{Switzerland}}}
\abstract{We investigate whether true physical observables associated with the measurements of large scale structure in the universe are frame-independent. 
In particular, we study if cosmological observables such as the galaxy number counts are invariant under the disformal transformations. In a previous work, it was shown that this frame-invariance holds true for the case of conformal transformations. In this work, we find that the invariance also holds true for the case of a simple disformal transformation.  We further briefly comment on the disformal invariance of other cosmological observables.}


\maketitle


\section{Introduction}
Scalar-Tensor (ST) theories are considered viable alternatives to the general theory of relativity (GR) and are usually at play while describing the inflationary epoch in the early universe or the present accelerated expansion in the late universe. While GR still remains the minimal theory of gravity which is remarkably successful in explaining diverse cosmological observations, there has been a long quest to explore various modifications of GR as well as to formulate viable alternatives and test them with multiple observations.  One of the well-known examples of a ST theory is the Brans-Dicke theory, in which the gravitational interaction is mediated by a scalar field in addition to the tensor field of GR \cite{Brans:1961sx} (See, for instance, Refs. \cite{Sotiriou:2008rp,Clifton:2011jh, Nojiri:2017ncd} for reviews on modified gravity). Moreover, there usually exists a non-trivial coupling between the gravitational sector and the matter sector in all such theories, and they also often have more than two propagating degrees of freedom \cite{Iyonaga:2021yfv}. In comparison, GR strictly has two such physical degrees of freedom, i.e. two polarization modes of the graviton. In order to be consistent with the observational tests on smaller scales such as the scales of our solar system, a realistic modification of gravity should necessarily contain a physical mechanism -- a screening mechanism to suppress the extra propagating scalar degrees of freedom on such scales \cite{DeFelice:2011th, Kimura:2011dc,Koivisto:2012za,Brax:2013ida,Brax:2021wcv}.

In order to construct viable modified theories, one must ensure that the additional degrees of freedom arising in such modifications do not lead to a ghost behaviour or other unwanted instabilities, both at the background and the perturbative order. These requirements led to the construction of the Horndeski theory -- the most general ST theory with a single scalar field leading to second-order equations of motion, thereby avoiding ghost degrees of freedom \cite{Horndeski:1974wa}, and possible ways to tackle instabilities regarding the speed of gravitational waves have also been discussed, for example in \cite{Oikonomou:2020sij}. These theories have found numerous applications in cosmology, particularly in constructing viable models of inflation and dark energy \cite{Deffayet:2013lga}.
Lately, healthy extensions beyond Horndeski theories have also been studied which are usually obtained by performing the so-called disformal transformation of the field space metric tensor which is essentially a generalization of the well-known conformal transformation (also sometimes called the Weyl transformation). Such generalisations can be useful in various ways, for example in the incorporation of quantum effects in spacetime geometry \cite{Chowdhury:2021fjw}, where disformal frames can help in dealing with singularity issues that arise in conformal frames, particularly because of the more general construction of a disformal transformation. A theory invariant under the conformal transformation is called conformally invariant which only involves rescaling of the metric tensor by a conformal factor. It is important to note that such a conformal transformation does not modify the causal structure of the spacetime. In general, the disformal transformation contains a scalar field $\phi$ as well as its first order derivatives\footnote{Note that the disformal transformation is not just a simple field
redefinition because it involves derivatives of the scalar field.} and can be written as 
\be 
g_{\mu \nu} = \mathcal{A}(\phi,X) \tilde{g}_{\mu \nu} + \mathcal{B}(\phi,X) {\nabla}_\mu \phi{\nabla}_\nu \phi\,, 
\label{e:disformal_def}
\ee 
where $\mathcal{A}$ and $\mathcal{B}$ are some general functions, referred to as the conformal and disformal factors, respectively and $X =-\frac{1}{2}g_{\mu\nu}{\nabla}^\mu \phi{\nabla}^\nu \phi = -\frac{1}{2}{\nabla}_\mu \phi{\nabla}^\mu \phi $.
Although ${\nabla}_\mu$ indicates the covariant derivative, for a scalar field it becomes ${\nabla}_\mu \phi = {\partial}_\mu \phi$.
For $\mathcal{B}=0$, Eq. (\ref{e:disformal_def}) reduces to the ordinary conformal transformation
while $\mathcal{A}=1$  represents the pure disformal transformation.
In \cite{Bekenstein:1992pj}, the concept and formulation of the disformal transformation was first introduced by Bekenstein in order to relate geometries of the same gravitational theory which suggests a richer set of possibilities for the transformed metric. Although Eq. (\ref{e:disformal_def}) does not indicate the most general form of a disformal transformation, it is one of the simplest ones involving the derivatives of a scalar. Since some of these possibilities may be unphysical, the functions $\mathcal{A}$ and $\mathcal{B}$ are generally subject to some constraints.

Over the years, disformal transformations have been applied in different contexts in GR and cosmology. 
For the case of massless Klein-Gordon equation \cite{Falciano:2011rf} and the vacuum Maxwell’s equations \cite{Goulart:2013laa}, the disformal transformations form a set of new symmetry transformations.  
Further, it has been shown that the gauge-invariant primordial cosmological perturbations are invariant under disformal transformation within the context of the Horndeski theory (in this context, see Refs. \cite{Minamitsuji:2014waa, Tsujikawa:2014uza, Domenech:2015hka, Motohashi:2015pra}). Moreover, they have been used to study black holes in the context of ST theory \cite{Babichev:2017guv}, to investigate allowed regions of the solution space and their symmetry \cite{BenAchour:2020wiw}, in the context of  dark energy \cite{Zumalacarregui:2010wj, Sakstein:2014isa, Brax:2015hma}, dark matter \cite{vandeBruck:2015ida,Brax:2020gqg} and more importantly, in generating a wider class of healthy ST theories (without additional number of physical degrees of freedom) beyond Horndeski theories, so called the DHOST theories \cite{Bettoni:2013diz, Gleyzes:2014qga,Gleyzes:2014dya,Zumalacarregui:2013pma, Langlois:2015cwa, BenAchour:2016cay,Crisostomi:2016tcp,Kobayashi:2019hrl}.

One may naively expect that true physical observables must be frame invariant. However, it may still be a tedious task to explicitly show this invariance  in different physical frames. Previously, some interesting studies on thermodynamic quantities \cite{Nashed:2019yto} and finite time singularity correspondence for various $F(R)$ gravity theories \cite{Bahamonde:2016wmz} have been carried out for Einstein and Jordan frames. It is well known that conformal transformations leave physical observables invariant. In an earlier paper \cite{Francfort:2019ynz}, we had shown that some specific observables associated with the distribution of large scale structure remain invariant under a purely conformal transformation. It was discussed that while the (unobservable) matter power spectrum is frame dependent, the observable number counts do not depend on the choice of the frame.
In this paper, we study whether these observables are invariant under the disformal transformations. As discussed in \cite{Bettoni:2013diz}, disformal transformations in the absence of the kinetic term represent a symmetry in the Horndeski action, analogous to conformal transformations and standard scalar-tensor theories, which indicates the invariance of physical observables. Since time and spatial coordinates are rescaled by the same factor under conformal transformations, they are usually referred to as the causality preserving transformations as the lightcone structure remains preserved. However, disformal transformations usually don't preserve causality as the time and spatial coordinates are rescaled by different factors. In order to ensure causal behaviour for all particles, one therefore requires the condition $\mathcal{B} <0$ everywhere. In fact, demanding that the disformally transformed metric is healthy and well defined everywhere, there arises a set of conditions that must be imposed on it: (i) it must be causal, (ii) it must preserve Lorentzian signature, (iii) the inverse must exist and be non-singular and (iv)  the volume element must be non-singular. All these conditions will ensure that the disformal frame is also a well-defined physical frame and the physical observables, if correctly calculated, must therefore be independent of the choice of any frame.
\par
Our approach in this paper is based on the seminal work of Bonvin and Durrer \cite{Bonvin:2011bg}, in which the authors derive a complete expression for cosmological number counts from fundamental principles, in terms of a sum of the density and volume perturbations. Since they already perform their calculations by solving the geodesic equation on the lightcone, we do not explicitly show these calculations for the disformal transformations, an approach that is usually undertaken in literature \cite{Sasaki:1987ad,Yoo:2009au,Yoo:2019qsl,Jeong:2011as,Schmidt:2012ne}. Instead, our approach is more simplified and treats the density and volume perturbations obtained in \cite{Bonvin:2011bg} as its backbone, using which we proceed with our calculations specific to the disformal transformations. Our aim is not to derive the individual contributors to the total number counts, but to emphasise on the starting point of this derivation, that is the contribution of the density and volume perturbations, and how they get modified under disformal transformations.

\vskip 4pt
\noindent
\textbf{Conventions and notations:}
We work with the $(-,+,+,+)$ signature. Various quantities in the Jordan frame are indicated with a tilde.

In a cosmological context, the FRLW metrics are

\begin{align}
    \mathbf{\tilde g} &=\tilde{a}^2( - \boldsymbol{\mathrm d}t^2 + \boldsymbol{\mathrm d }\boldsymbol{x}^2),, \\
    \mathbf{g} &= \tilde{a}^2(-\alphaT^2  \boldsymbol{\mathrm d}t^2 + \alphaL^2  \boldsymbol{\mathrm d }\boldsymbol{x}^2)\,,
\end{align}
where $t$ denotes conformal time.
with
\be 
\alphaT^2 = \mathcal A - \mathcal B \dot{\phi}^2\,, \quad
\alphaL^2 = \mathcal A \,.
\ee 
This implies that the measurements of length and time in both frames are related as
\be  \label{eq:intro_LT}
L = \alphaL \tilde L \,, \quad 
T = \alphaT \tilde T\,.
\ee

This paper is organized as follows.
In Section \ref{s:background_perturb}, we point out the scaling of various relevant quantities, including the background variables and first order perturbations, that will be useful for our calculations.
In Section \ref{s:num_counts}, we obtain the  transformations of the density and volume perturbations, and prove that the cosmological number counts are indeed equal in the Jordan and Einstein frames. Finally, in Section \ref{s:conclusion} we summarise our results and provide an outlook for future directions.

\section{Background and perturbed quantities}
\label{s:background_perturb}
In this section, we shall study the transformations of various background and perturbed quantities under the disformal transformations which, as in Eq. (\ref{e:disformal_def}), is usually written as
$$
\,g_{\mu \nu} = \mathcal{A}(\phi,X) \tilde{g}_{\mu \nu} + \mathcal{B}(\phi,X) {\nabla}_\mu \phi{\nabla}_\nu \phi\,, \quad
\mathrm{with} \quad X =-\frac{1}{2}g_{\mu\nu}{\nabla}^\mu \phi{\nabla}^\nu \phi = -\frac{1}{2}{\nabla}_\mu \phi{\nabla}^\mu \phi\,.
$$ 
Note that, as $\phi$ is a scalar, the covariant derivative is nothing but a partial derivative, hence we do not put any tilde to avoid heavy notations. We also drop the $X$ dependence of $\mathcal{A}$ and $\mathcal{B}$ which has been discussed in Section \ref{ss:backvar}.
\subsection{Background variables}\label{ss:backvar}
We first proceed by obtaining the relations between the background variables in the Jordan and Einstein frames, followed by that of the perturbations as well in these two frames. The quantities that we are primarily interested in this section are the Hubble parameter $\mathcal{H}$, the energy density $\rho$ and the pressure $P$. In case of the disformal transformation, we can get the relation between the comoving Hubble parameters in the two frames as follows:
\begin{eqnarray}
 \tilde{\mathcal{H}}&=&\frac{\dot{\tilde{a}}}{\tilde a}=\frac{\mathrm{d}}{\mathrm{d}t}\left(\frac{a}{\alpha_{0L}}\right)\frac{\alpha_{0L}}{a}\nonumber\\
  &=&\mathcal{H}-\frac{\alpha_{0L}^\prime}{\alpha_{0L}}\dot{\phi}_0=\mathcal{H}\frac{\dot L_0}{L_0}
\end{eqnarray}
Here, the quantities with subscript $0$ indicate the background quantities, a dot denotes derivative with respect to $t$, and a prime denotes derivative with respect to $\phi_0$. It is important to note here that we are expressing the $\alpha$'s in terms of $L$'s by considering the standard ruler in the Jordan frame to be unity, that is to say, $L=\alpha_L\tilde L=\alpha_L$. 
 From the energy-momentum tensors in the two frames, given by
\begin{eqnarray}
T^{\mu}{ }_{\nu}&=&\rho u^{\mu} u_{\nu}+P\left(u^{\mu} u_{\nu}+\delta^{\mu}{ }_{\nu}\right), \\
\tilde{T}^{\mu}{ }_{\nu}&=&\tilde{\rho} \tilde{u}^{\mu} \tilde{u}_{\nu}+\tilde{P}\left(\tilde{u}^{\mu} \tilde{u}_{\nu}+\delta^{\mu}{ }_{\nu}\right),
\end{eqnarray}
we can obtain a generalised relation between the energy-momentum tensor components, i.e., the energy density and the pressure in the two frames in terms of the kinetic term $X$ as \cite{Chiba:2020mte}
\begin{eqnarray}
\rho&=&\xi_1\tilde{\rho}+\xi_2\tilde{P}\label{e:rho_dis}\\
P&=&\frac{\tilde{P}}{\alpha\mathcal{A}^{3/2}}=\frac{\tilde{P}}{\alpha_T\alpha_L^3}
\label{e:P_dis}
\end{eqnarray}
where $\xi_1$ and $\xi_2$ are given by
\begin{eqnarray}
\xi_{1}&=&\frac{(1-2 \mathcal{B} X / \mathcal{A})^{1 / 2}}{\mathcal{A}\left(\mathcal{A}-\mathcal{A}_{, X} X+2 \mathcal{B}_{, X} X^{2}\right)}
\label{e:xi1_chiba}
\\
\xi_{2}&=&-\frac{3(1-2 \mathcal{B} X / \mathcal{A})^{1 / 2} \mathcal{A}_{, X} X}{\mathcal{A}^{2}\left(\mathcal{A}-\mathcal{A}_{,X} X+2 \mathcal{B}_{, X} X^{2}\right)}
\label{e:xi2_chiba}
\end{eqnarray}
where $\mathcal{A}_{,X}=\partial\mathcal{A}/\partial X$.
However, the dependence on $X$ is often done away with because it introduces a metric dependence in the conformal factor which brings extra terms that the standard scalar-tensor theory fails to accommodate. Also, if we wish to extend the formalism to Horndeski theory, for example, the Horndeski structure is seen to be invariant only when the $X$ dependence is dropped \cite{Zumalacarregui:2012us,Bettoni:2013diz}. Moreover, for our purpose, the computation of number counts requires only the density and volume perturbations, so we can neglect the pressure term above.
Then we find 
\be
    \xi_1=\alpha\mathcal{A}^{-5/2}=\frac{\alpha_T}{\alpha_L^5}\quad\text{and}\quad
    \xi_2 =0
\ee
These equations simply follow from the fact that $\alpha_T=\sqrt{\mathcal{A}-\mathcal{B}\dot\phi^2}=\sqrt{\mathcal{A}-2\mathcal{B}X}$, since $2X=-{\nabla}_\mu \phi{\nabla}^\mu \phi$.\par
The conservation of the energy-momentum tensor $\tilde\nabla^\mu \tilde{T}_{\mu\nu}=0$ in the Jordan frame gives
\begin{equation}
    \dot{\tilde{\rho}}=-(\tilde{\rho}+\tilde P)3\tilde{\mathcal{H}}
\label{e:conservation_jordan}
\end{equation}
So, from Eq. (\ref{e:rho_dis}) and the expression of $\xi_1$ above, we have,
\begin{eqnarray}
    \dot{\rho}&=&\dot{\xi}_1\tilde{\rho}+\xi_1\dot{\tilde{\rho}}\nonumber\\
    &=& -3\mathcal{H}\rho-\frac{2\alpha_L^\prime\dot{\phi}}{\alpha_L}\rho+\frac{\alpha_T^\prime\dot{\phi}}{\alpha_T}\rho
\label{e:rhodot}
\end{eqnarray}
In the conformal case, $\xi_1=\mathcal{A}^{-2}$, $\xi_2=0$ and $\alpha_L=\sqrt{\mathcal{A}}$.\par
\subsection{First order perturbations}
We now turn to the first order perturbed quantities for scalar perturbations and  neglect the vector and tensor perturbations. In what follows, the perturbation in any quantity is denoted with a $\delta$, for example, $\delta\phi$ corresponds to a perturbation in $\phi$. The expression for a perturbed $\xi_1$ turns out to be
\begin{eqnarray}
    \xi_1 &=& \xi_{10}\left(1-\frac{5}{2}\frac{\mathcal{A}_0^\prime\delta\phi}{\mathcal{A}_0}+\frac{\alpha^\prime_{0T}\delta\phi}{\alpha_{0T}}\right)\nonumber\\
    &=& \xi_{10}\left(1-\frac{5\alpha_{0L}^\prime\delta\phi}{\alpha_{0L}}+\frac{\alpha_{0T}^\prime\delta\phi}{\alpha_{0T}}\right)
\end{eqnarray}
where $\xi_{10}$ is the background quantity.
Hence, the usual gauge-dependent density fluctuations in both the frames can be related as,
\begin{eqnarray}
    \delta(x)&=& \frac{\rho(x)-\rho_0(t)}{\rho_0(t)}=\frac{\xi_1\tilde{\rho}(x)-\xi_{10}\tilde{\rho}_0(t)}{\xi_{10}\tilde{\rho}_0(t)}\nonumber\\
    &=&\tilde{\delta}(x)-\left(\frac{5\alpha_{0L}^\prime\delta\phi}{\alpha_{0L}}-\frac{\alpha_{0T}^\prime\delta\phi}{\alpha_{0T}}\right)\nonumber\\
     &=&\tilde{\delta}(x)-\left(\frac{5\delta L}{L}-\frac{\delta T}{T}\right)
\label{e:deltax_dis}
\end{eqnarray}
We will later be focusing on gauge-invariant density perturbations on the lightcone, which are a function of the redshift and direction only, and which can help us relate directly to physical observables.

\section{Cosmological number counts}\label{s:num_counts}
The cosmological number counts are observables that are computed in terms of the number $N$ of galaxies observed in a patch of the sky in a particular direction $\mathbf{n}$ per unit solid angle and per unit redshift bin and are quantified as the perturbation in the number density of galaxies as \cite{Bonvin:2011bg}
\begin{equation}
  \Delta(\mathbf{n}, z)=\frac{N(\mathbf{n}, z)-N_0(z)}{N_0(z)}
\label{e:numcount_rhs}
\end{equation}
where $N_0$ is the number of galaxies averaged over angles. It can be seen that the right hand side of Eq. (\ref{e:numcount_rhs}) follows directly from the definition of the redshift density perturbation (Eq. (1) of \cite{Bonvin:2011bg}):
\begin{eqnarray}
\tilde\delta_{z}(\mathbf{n}, z) &=&\frac{\tilde\rho(\mathbf{n}, z)-\tilde\rho_{0}(z)}{\tilde\rho_{0}(z)}
= \frac{\frac{\tilde N(\mathbf{n}, z)}{\tilde V(\mathbf{n}, z)}-\frac{\tilde N_{0}(z)}{\tilde V(z)}}{\frac{\tilde N_{0}(z)}{\tilde V(z)}}\nonumber\\
&=&\tilde \Delta(\mathbf{n}, z)-\frac{\delta\tilde  V(\mathbf{n}, z)}{\tilde V(z)}
\label{e:dens_perturb_fundamental}
\end{eqnarray}
where we have used tilde to denote quantities in the Jordan frame. Hence,
\begin{equation}
\tilde{\Delta}(\mathbf{n}, z)=\tilde{\delta}_{z}(\mathbf{n}, z)+\frac{\delta \tilde{V}(\mathbf{n}, z)}{\tilde{V}(z)}
\label{e:num_counts_jordan}
\end{equation}
As mentioned earlier, the first term $\tilde{\delta}_z$ is the redshift-space density perturbation, and the second term is the volume perturbation divided by the physical survey volume density per
redshift bin, per solid angle. Both these quantities are gauge-invariant, but not frame-invariant observables on their own accord. Our aim is to show that their sum $\tilde{\Delta}(\mathbf{n}, z)$ is invariant under a disformal transformation and is the same as the quantity in the Einstein frame, which we will denote without a tilde, $\Delta(\mathbf{n}, z)$.\par
From the fundamental definition of the density perturbation in the redshift space, we have
\begin{eqnarray}
\delta_{z}(\mathbf{n}, z) &=&\frac{\rho(\mathbf{n}, z)-\rho_{0}(z)}{\rho_{0}(z)}\nonumber\\
&=& \frac{\frac{N(\mathbf{n}, z) (\alpha_T/\alpha_L^2)^{-1}}{V(\mathbf{n}, z)}-\frac{N_{0}(z) (\alpha_{0T}/\alpha_{0L}^2)^{-1}}{V(z)}}{\frac{N_{0}(z) (\alpha_{0T}/\alpha_{0L}^2)^{-1}}{V(z)}}\nonumber\\
&=&\Delta(\mathbf{n}, z)-\frac{\delta V}{V}+2\left(\delta z\frac{1}{L_0} \frac{\dd L_0}{\dd z} - \frac{\delta L_0}{L_0}\right)-\left(\delta z\frac{1}{T_0} \frac{\dd T_0}{\dd z} - \frac{\delta T_0}{T_0}\right)
\label{e:dens_perturb_fundamental}
\end{eqnarray}
The second equality follows from the fact that $\rho=mN/V$ and the mass $m$ scales as $\alpha_T/\alpha_L^2$. The last equality is self-explanatory from the definition given in Eq \ref{e:num_counts_jordan} and the fact that $L(\mathbf{n}, z)=L\left(z_{0}\right)+\delta L(\mathbf{n}, z)$. The quantity $\delta z$ corresponds to the splitting of the observable redshift $z$ into a background and a perturbative part, that is, $z=z_0+\delta z$.\par
For a general function $f$, we can assume that it has time and length dimensions $n_T$ and $n_L$ respectively, and the generalization of the formula becomes
\be 
f = \tilde f L^{n_L}T^{n_T},
\ee
from which we obtain after some basic calculation (see \cite{Francfort:2019ynz} for details),
\be
\frac{\delta f (\boldsymbol n,z)}{f(z)} =\frac{\delta \tilde f(\boldsymbol n,z)}{\tilde f(z)} +\left(
n_L \frac{\delta z}{L_{0}} \frac{\mathrm{d} L_{0}}{\mathrm{~d} z_{0}}-n_L \frac{\delta L}{L}\right)+
\left(n_T \frac{\delta z}{T_{0}} \frac{\mathrm{d} T_{0}}{\mathrm{~d} z_{0}}-n_T \frac{\delta T}{T}\right)\,,
\label{e:general_formula}
\ee
where $\delta f$ is the first order perturbation in the function $f$, as a function of direction $\mathbf{n}$ and redshift $z$ and calculated in the Einstein frame and $\delta \tilde f$ is the corresponding quantity in the Jordan frame. 
\subsection{Density perturbations}


In this section, we aim to obtain a relation for the density perturbations between the two frames. Following the definition of the redshift-space density perturbation in \cite{Bonvin:2011bg}, we have
\bea
    \delta_{z}(\mathbf{n}, z) &=&\frac{\rho(\mathbf{n},z)-\rho_0(z)}{\rho_0(z)}\nonumber\\
    &=&\frac{\delta \rho(\mathbf{n}, z)}{\rho_0\left(z_{0}\right)}-\frac{\mathrm{d} \rho_{0}}{\mathrm{~d} z_{0}} \frac{\delta z(\mathbf{n}, z)}{\rho_{0}(z_{0})}
\label{e:redshiftspace_densperturb}
\eea
where the first term corresponds to the relation established by Eq. (\ref{e:deltax_dis}). For the second term we first need to compute the expressions for $\mathrm{d}z_0/dt$ and $\mathrm{d}\rho_0/dt$.
The latter has already been done in Eq. (\ref{e:rhodot}). In the Einstein frame we have,
\begin{equation}
    1+z_0=\frac{\alpha_{0L}}{a}
\end{equation}
which implies that 
    \begin{equation}
    \frac{dz_0}{dt}=-\frac{\dot{a}}{a^2}\alpha_{0L}+\frac{\dot{\alpha_{0L}}}{a}=-(1+z_0)\left(\mathcal{H}-\frac{\dot L_0}{L_0}\right)
\label{e:zdot_dis}
\end{equation}
We can now use this expression to obtain
\begin{eqnarray}
    \frac{\mathrm{d}\rho_0}{\mathrm{d}z_0}&=&\frac{\mathrm{d}\rho_0}{\mathrm{d}t}\frac{\mathrm{d}t}{\mathrm{d}z_0}\nonumber\\
    &=& \frac{-3\mathcal{H}\rho_0-\frac{2\dot L_0}{L_0}\rho_0+\frac{\dot T_0}{T_0}\rho_0}{-(1+z_0)\left(\mathcal{H}-\frac{\dot L_0}{L_0}\right)}
\end{eqnarray}
Now, from Eq. (\ref{e:redshiftspace_densperturb}), and realising that the first term on the right hand side is essentially the quantity $\delta(x)$ as defined in Eq. (\ref{e:deltax_dis}), we get an expression for the density contrast $\tilde{ \delta}_z$ in the Jordan frame as
\begin{equation}
   \tilde{ \delta}_z(\mathbf{n},z) = \tilde{\delta}(x) - \left(\frac{\mathrm{d}\rho_0}{\mathrm{d}z_0}\right)\frac{\delta z (\mathbf{n},z)}{\rho_0(z_0)}
   \end{equation}
   which, using the previous equation, yields
   \begin{equation}
    \tilde{\delta}(x)= \tilde{\delta}_z-\frac{3\mathcal{H}+\frac{2\dot L_0}{L_0}-\frac{\dot T_0}{T_0}}{(1+z_0)\left(\mathcal{H}-\frac{\dot L_0}{L_0}\right)}\delta z
\end{equation}
This relation helps us transform the gauge-dependent density fluctuations $\tilde\delta(x)$ into the gauge-invariant density perturbations $\tilde\delta_z$.
Upon using Eq. (\ref{e:deltax_dis}), we find 
\begin{equation}
    \delta(x)=\tilde\delta_z-\frac{3\mathcal{H}+\frac{2\dot L_0}{L_0}-\frac{\dot T_0}{T_0}}{(1+z_0)\left(\mathcal{H}-\frac{\dot L_0}{L_0}\right)}\delta z-\left(\frac{5\delta L}{L}-\frac{\delta T}{T}\right)
\end{equation}
which leads to 
\begin{eqnarray}
    \delta_z &=& \tilde{\delta}_z-\left(\frac{5\delta L}{L}-\frac{\delta T}{T}\right)\nonumber\\
     &=& \tilde{\delta}_z-5\frac{\delta L}{L_0}+5\frac{\mathrm{d}L_0}{L_0\mathrm{d}z_0}\delta z+\frac{\delta T}{T_0}-\frac{\mathrm{d}T_0}{T_0\mathrm{d}z_0}\delta z
\label{e:dens_perturb}
\end{eqnarray}
The relation in the above equation gives us the ``unobservable" density contrast $\delta_z$ in the Einstein frame in terms of the same quantity in the Jordan frame, and as is evident, they are unequal. This renders the quantity $\delta_z$ frame-dependent, although it is gauge-invariant, as already discussed. This also automatically implies that the ``unobservable" matter power spectrum $P_{\delta_z}(k,z)$ which is nothing but a Fourier transform of the two-point correlation function of the density contrast, is also frame-dependent, as can be seen from the following definition of the matter power spectrum
\begin{equation}
    \langle\delta_z(\mathbf{k})\delta_z^*(\mathbf{k'})\rangle=(2\pi)^3 P_{\delta_z}(k,z)\delta^3(\mathbf{k-k'})\neq (2\pi)^3 P_{\tilde\delta_z}(k,z)\delta^3(\mathbf{k-k'})
\label{e:powspec}
\end{equation}
where $\delta^3$ is the three-dimensional Kronecker delta function. Contrary to this result, in the next section, we will show that the ``observable" matter power spectrum that includes both the density and volume perturbations is indeed frame-invariant.
\subsection{Volume perturbations}
Since the volume in the Einstein frame scales as $\alpha_L^3$, we can make use of the generalised relation in Eq. (\ref{e:general_formula}) to obtain the relation between the volume perturbations in the Einstein and Jordan frames.
\begin{eqnarray}
   \frac{\delta V}{V} &=& \frac{\delta \tilde{V}}{\tilde{V}}+3 \frac{\alpha_L^\prime\dot{\phi}}{\alpha_L}=\frac{\delta\tilde V}{\tilde V}+3 \frac{\delta L}{L}\nonumber\\
    &=& \frac{\delta \tilde{V}}{\tilde{V}}+3\frac{\delta L}{L_0}-3\frac{\mathrm{d}L_0}{L_0\mathrm{d}z_0}\delta z
\label{e:vol_perturb}
\end{eqnarray}
From Eqs. (\ref{e:dens_perturb_fundamental}), (\ref{e:dens_perturb}) and (\ref{e:vol_perturb}) we finally have
\begin{eqnarray}
\Delta_E=\Delta (\mathbf{n},z) &=& \delta_z+\frac{\delta V}{V}-2\left(\delta z\frac{1}{L_0} \frac{\dd L_0}{\dd z_0} - \frac{\delta L}{L_0}\right)+\left(\delta z\frac{1}{T_0} \frac{\dd T_0}{\dd z_0} - \frac{\delta T}{T_0}\right)\nonumber\\
&=& \left(\tilde{\delta}_z-5\frac{\delta L}{L_0}+5\frac{\mathrm{d}L_0}{L_0\mathrm{d}z_0}\delta z+\frac{\delta T}{T_0}-\frac{\mathrm{d}T_0}{T_0\mathrm{d}z_0}\delta z\right)+\left(\frac{\delta \tilde{V}}{\tilde{V}_{0}}+3\frac{\delta L}{L_0}-3\frac{\mathrm{d}L_0}{L_0\mathrm{d}z_0}\delta z\right)\nonumber\nonumber\\
&& -2\left(\delta z\frac{1}{L_0} \frac{\dd L_0}{\dd z_0} - \frac{\delta L}{L_0}\right)+\left(\delta z\frac{1}{T_0} \frac{\dd T_0}{\dd z_0} - \frac{\delta T}{T_0}\right)\nonumber\\
&=& \tilde{\Delta} (\mathbf{n},z) = \Delta_J
\end{eqnarray}
where the subscripts $E$ and $J$ stand for Einstein and Jordan frames, respectively.

Therefore, we can show that in this case the matter power spectrum in both frames would indeed be invariant 
\begin{equation}
    \langle\Delta(\mathbf{k})\Delta^*(\mathbf{k'})\rangle=    \langle\tilde\Delta(\mathbf{k})\tilde\Delta^*(\mathbf{k'})\rangle\nonumber
    \end{equation}
    Or, equivalently, 
    \begin{equation} (2\pi)^3 P_\Delta(k,z)\delta^3(\mathbf{k-k'})=(2\pi)^3 P_{\tilde\Delta}(k,z)\delta^3(\mathbf{k-k'})
\end{equation}
This establishes an interesting and important fact that while the ``unobservable" power spectrum of the density contrast is frame-dependent, the ``observable" matter power spectrum is indeed frame-invariant.
\section{Conclusions}\label{s:conclusion}
Disformal transformations are considered a generalisation of the conformal transformations which also involve derivative-dependent  terms of a scalar/vector field. 
While the conformal transformations usually involve a rescaling of the metric which also preserve causality, the disformal transformations are more general which affect the particle geodesics and also lead to a non-trivial coupling of the matter Lagrangian. These salient features provide very rich phenomenology which has been explored in different cosmological contexts, in black hole physics and more interestingly, in constructing an even more general class of ST theories beyond Horndeski theories. 

In this paper, we have investigated how physical observables associated with the galaxy surveys in the Jordan and
Einstein frames are related by a simple disformal transformation constructed by scalar fields, taking into consideration that the disformal frames are well-defined and free from acausality. In particular, we have successfully been able to show the  frame-invariance of the cosmological number counts for the case of such a disformal transformation which is an interesting result, since it represents a symmetry of the Horndeski action. This establishes the property of number counts as a physical observable even further. A summary of our results that we have obtained in this work, has been outlined in Table \ref{t:summary}, where we can see that the background density, pressure and redshift are not gauge-invariant, and hence not observable. The same holds true for the perturbed density and velocity, since we understand from \cite{Bonvin:2011bg} that only the corresponding quantities in the redshift space are gauge-invariant. It is to be noted that frame-invariance is crucial for physical observability on the telescope, while gauge-invariance is essentially a theoretical construct. Our results thus establish that the truly observable quantity is the number counts $\Delta(\mathbf{n},z)$ which are both gauge and frame-invariant.
\begin{table*}[t]
\begin{center}                      
    \begin{tabular}{|p{3cm}|p{2.6cm}|p{1.1    cm}|p{5cm}|}
    \hline  & & \multicolumn{2}{l|}{}\\
    Quantity            & Gauge-invariant & \multicolumn{2}{c|}{Frame-invariant~~~~~~} \\  \hline
    {\bf Background} &&& \\
    Density $\rho_0$ & No  &No & $\rho_0 = \frac{\alpha_{0T}}{\alpha_{0L}^5}\tilde \rho_0$\\
    Pressure $P_0$ & No & No & $P_0 = \frac{\alpha_{0T}}{\alpha_{0L}^3}\tilde P_0$\\
    Redshift $z_0$  & No & Yes & \\
    Observed redshift $z$   & Yes & Yes & \\ \hline
    {\bf Perturbations} &&& \\
    Density $\delta$ & No & No &  $\delta=\tilde\de -\left(\frac{5\delta L}{L}-\frac{\delta T}{T}\right)$ \\
    Velocity $\bv$ & No & Yes & \\
   Redshift density $\delta_z$ & Yes & No & $\delta_z=\tilde{\delta}_z-5\frac{\delta L}{L_0}+5\frac{\mathrm{d}L_0}{L_0\mathrm{d}z_0}\delta z+\frac{\delta T}{T_0}-\frac{\mathrm{d}T_0}{T_0\mathrm{d}z_0}\delta z$ \\
    Volume perturbation $\frac{\delta V}{V}$ & Yes & No & $\de V/V=\frac{\delta \tilde{V}}{\tilde{V}_{0}}+3\frac{\delta L}{L_0}-3\frac{\mathrm{d}L_0}{L_0\mathrm{d}z_0}\delta z$\\
    Number counts $\De(\bn,z)$ & Yes & \textbf{Yes} &\\  
    \hline
    \end{tabular}
    \end{center}
    \vspace{6pt}
    \caption{Gauge and frame invariance of various quantities.  ~} 
    \label{t:summary}
\end{table*}
\par Although we have only considered the disformal factor $\cal B$ as a function of $\phi$ only, and it might also seem non-trivial to carry out the same kind of proof for a disformal transformation including the $X$ dependence, we believe that our calculations provide a first step towards a necessary generalisation, as we also observe that our results reduce to those of \cite{Francfort:2019ynz} under the pure conformal limit. However, an apparent appearance of Ostrogradsky ghosts
for a $X$ dependent disformal transformation might happen, no real ghost in the theory should actually be present in the transformed frame due to the existence of
hidden constraints \cite{Zumalacarregui:2013pma}. 
Our work can, in fact, be further extended to more complex transformations, for example to those in \cite{Domenech:2015hka,Motohashi:2015pra,Tsujikawa:2014uza,Alinea:2020laa}, employing much more rigorous and non-trivial calculations, which is beyond the scope of the current work.

Another direction in which one can explore the frame-invariance of cosmological number counts is by further generalisation of the Brans-Dicke ST theory itself, via the Horndeski Lagrangian \cite{Horndeski:1974wa}. An attempt to establish the invariance of the Horndeski Lagrangian under disformal transformations has been made in \cite{Bettoni:2013diz}, where it was found that the $X$ dependence is detrimental to the frame invariance. However, it would be interesting to explore how the frame-invariance of number counts holds up beyond the ST theories.
We leave these interesting directions for future work.
\section*{Data availability statement}
There is no new data generated or associated with this work.
\section*{Acknowledgments}
We would like to thank Ruth Durrer for very useful discussions and comments on the draft. BG acknowledges partial financial support from the CV Raman Postdoctoral Fellowship and the DST-INSPIRE Faculty Fellowship DST/INSPIRE/04/2020/001534. JF acknowledges financial support from the Swiss National Science Foundation. RKJ wishes to acknowledge financial support from the new faculty seed start-up grant of the Indian Institute of Science, Bengaluru, India, Science and Engineering Research Board, Department of Science and Technology, Government of India, through the Core Research Grant~CRG/2018/002200 and the Infosys Foundation, Bengaluru, India through the Infosys Young Investigator award.

\bibliography{disformal}

\end{document}